*************************************************************************
 *  All preprint and All figures for the preprints                       *
 *  from Nuclear Physics Group of                                        *
 *  Drexel University are available through ftp approach.                *
 *  The procedure is : ftp at                                            *
 *        einstein.physics.drexel.edu                                    *
 *        anonymous                          (as username)               *
 *        <Return>                           (as password)               *
 *        cd pan_files                       (go to the subdirectory)    *
 *************************************************************************

 For the paper entitled
      Single-Particle Nondegeneracy and SU(3) Fermion Dynamical Symmetry:
 The filenames for the one figures are split.fig1

\documentstyle[preprint,aps]{revtex}
\begin{document}
\baselineskip 0.75cm

\begin{center}
\ \\
 {\LARGE{\bf Single-Particle Nondegeneracy and SU(3) \\
Fermion Dynamical Symmetry \\ }}
\  \\
\vspace{0.5cm}
\  \\ {\large Cheng-Li Wu$^{1}$,  Da Hsuan Feng$^{2}$, Mike W. Guidry$^{3,4}$,
\\[2pt] Hsi-Tseng Chen$^{1}$, and Xing-Wang Pan$^{2}$}
\\
\vspace{3pt}
{\vbox{
     \small\em
      $^{1}$Department of Physics, Chung Yuan Christian University,
     Chung-Li, Taiwan 32023 ROC }}

\vspace{3pt}
{\vbox{
     \small\em
     $^{2}$Department of Physics and Atmospheric Sciences, \\
      Drexel University, Philadelphia, PA 19104 USA}}

\vspace{3pt}
{\vbox{
     \small\em
     $^{3}$Department of Physics, University of Tennessee, Knoxville, TN
     37996--1200 USA}}

\vspace{3pt}
{\vbox{
     \small\em
      $^{4}$Physics Division, Oak Ridge National Laboratory
      \small\em
      Oak Ridge, TN 37831 USA}}

\ \\

\end{center}
\vfill
\begin{center}
\vspace{0.5cm} {\large{\bf Abstract}}
\end{center}

\noindent{\small It is shown that the $SU(3)$ symmetry of the fermion dynamical
symmetry model is essentially preserved even for highly nondegenerate spherical
single-particle energies.  The breaking of $SU(3)$ symmetry by single-particle
energy terms for either normal deformation or superdeformation occurs only
through an indirect Pauli effect and is significant only when the spherical
single-particle splitting within shells is artificially large relative to that
observed experimentally.}


\newpage

The shell model is commonly accepted as the microscopic basis for nuclear
structure, but its practical implementation in medium and heavy nuclei requires
a severe truncation of the model space.  In the late 1980's, motivated by the
phenomenological successes of the Interacting Boson Model \cite{iac87} and
building on the schematic fermion model of Ginocchio \cite{gin80}, we proposed
a symmetry-dictated  truncation scheme for the shell model termed the Fermion
Dynamical Symmetry Model ({\small FDSM}) \cite{wu86,wu87b}. The symmetry limits
of the model and perturbations around these limits have been explored
extensively, and found to be consistent with a broad range of nuclear structure
observations \cite{wu93b}.  It is now of interest to examine in detail the
excursions from the symmetry limits of the theory, in order to test its
suitability as a  systematic truncation procedure for quantitative shell model
calculations in heavy nuclei. Although symmetry breaking has been investigated
in some cases,  there is as yet no systematic analysis of such terms in the
theory.  In this paper, we initiate such an analysis for the $Sp(6) \supset
SU(3)$  limit of the {\small FDSM}.

The $Sp(6) \supset SU(3)$ dynamical symmetry of the {\small FDSM} may be
identified with axially-symmetric rotational motion, and matrix elements
derived in the symmetry limit or in perturbation around this limit have been
shown to be in quantitative agreement with a variety of collective observables
in heavy rotational nuclei \cite{wu93b}.  The {\small FDSM}  uses a modified
Ginocchio coupling scheme that decomposes the single-particle angular
momenta $j$ of the shell model into an integer part $k$ and a half-integer part
$i$ such that $\bf j = \bf k + \bf i$.  For orbitals exhibiting an $Sp(6)
\supset SU(3)$ dynamical symmetry, $k=1$.  In the lighter nuclei, there is a
single value of $i$ within a shell.  In heavier nuclei there are typically 2--3
values of $i$ within the normal-parity orbitals of a major shell, and for the
enlarged valence spaces characteristic of superdeformation there may be as many
as 5 values of $i$ within a supershell \cite{wu92b}.  A Hamiltonian with an
$Sp(6) \supset SU(3)$ dynamical symmetry requires that single-particle energy
terms corresponding to the same value of $i$ be degenerate.  Thus, the
symmetry-limit Hamiltonian will generally exhibit a higher level of degeneracy
than the realistic spherical single-particle shell model spectrum, and
quantitative calculations must consider the effect on the symmetry-limit
results of symmetry-breaking by single-particle energies. We emphasize that
these remarks concern the splitting of the single-particle spectrum for the
{\em spherical} shell model. The additional splitting associated with
quadrupole interactions (the algebraic analog of Nilsson splittings at finite
deformation) is a separate issue that is handled in the {\small FDSM} through
quadrupole--quadrupole coupling terms of the 2-body Hamiltonian.  The major
portion of these terms respects the symmetry \cite{wu93b}, and the remainder
may be incorporated numerically where needed \cite{hwu87}.

In the discussion of the single-particle splitting, the question of the
physically relevant scale for the phenomenon is important. As has been
discussed extensively in Refs.\ \cite{wu92b,wu93b,gui93d}, for collective
properties of many-body systems a natural scale is set by the dominant
correlation energies. The issue that must be addressed is not simply the
size of the single-particle splitting, but its size relative to the correlation
energy of the system and {\em how much of that splitting breaks the relevant
symmetry.}  In particular, no matter how large the single-particle splitting
terms are,  they will have no influence on the $SU(3)$ properties such as
moments of inertia if they commute with the invariants of $SU(3)$. Furthermore,
even if they do not commute, their influence will be greatly suppressed if
correlation energies in the system produce large energy separations between
irreducible representations of the dynamical symmetry.

To begin, we  rewrite the single-particle energy in terms of the standard
{\small FDSM} $k$--$i$ basis:
\begin{equation} \sum_{j} n_{j}e_{j}=\sum_{r, i} { n}^{(rr)0}_{i}
e^{r}_{i},
\label{1}
\end{equation}
\begin{equation} { n}^{(rr)0}_{i} =\sqrt{2\Omega_{i}} \left[
b_{ki}^{\dagger} \tilde{b}^{\phantom{\dagger}}_{ki} \right]^{(rr)0} \qquad
e^{r}_{i}=\sum_{j}e_{j}
\left[ \begin{array}{ccc}
k&i&j\\
k&i&j\\
r&r&0 \end{array} \right]
\sqrt {\Omega_{j}
/ \Omega_{i}}.
\label{3}
\end{equation}
with the square bracket denoting a normalized 9-j  coefficient.  The
$k$--$i$ basis
$b_{ki}^{\dagger}$ has been defined in
\cite{gin80,wu86,wu87b}; $\Omega_{j}$ and $\Omega_{i}$ are the pair
degeneracies for the
$j$ shell and the shells associated with pseudospin $i$, respectively
[$\Omega_{j}=j+\frac{1}{2}$, and $\Omega_{i}=(2k+1)(2i+1)/2$].

The states of the {\small FDSM} are classified according to a total heritage
quantum number $u$ that measures the number of particles not coupled to
coherent S and D pairs \cite{wu87b}.  States below the first backbending
region are dominantly
$u=0$ configurations. The mixing matrix elements associated with the splitting
of the single-particle energies may be expressed as
\begin{equation}
\left< \lambda' \mu' u' \left| n_i^{(rr)0} \right| \lambda \mu u \right> =
\frac{\left< \lambda' \mu' u' \left| [n_i^{(rr)0}, C_{SU(3)}] \right|
\lambda \mu u \right>} {C(\lambda\mu) - C(\lambda'\mu')}.
\end{equation} where $u$ and $u'$ are the heritage quantum numbers and
$C(\lambda \mu)$ is the usual eigenvalue of the quadratic $SU(3)$ Casimir
operator $C_{SU(3)}$ evaluated in an $SU_{3}$ representation $(\lambda, \mu)$.
Therefore, we must  examine the commutation of the operators  ${
n}^{(rr)0}_{i}$ for $r = 0,1, 2$ with the {\small FDSM} $SU(3)$ Casimir
operator
$C_{SU(3)}$.
After some extensive algebra, one finds that ${ n}^{(11)0}_{i}$ is the only
component of the single-particle operator that can mix an $SU(3)$ irrep in the
$u=0$ bands  with other $SU(3)$ irreps. The resulting
mixing matrix element may be expressed in the form
\begin{eqnarray}
 & & \hspace{-0.5 in} \left<\lambda'\mu' u' \left| \sum_{r, i} {
n}^{(rr)0}_{i} e^{r}_{i} \right| \lambda
\mu u=0 \right>
                                    \nonumber \\
 & & \hspace{-0.5 in} = \delta_{r1} \delta_{u'2}
\left\{ \ -4
\sum_{i}e^{1}_{i}
\left[
\begin{array}{ccc} i&i&1\\ i&i&0\\ 1&0&1
\end{array}
\right]
 \frac{\left<\lambda'\mu' u=2 \left| [{\cal A}_{11}^\dagger(i)
\tilde{\cal A}_{00}(i)]^{(11)0} \right| \lambda \mu u=0
\right>}{\Delta C}
\right.\nonumber \\[5pt]
 & &
 \left. + 4.472
\sum_{i}e^{1}_{i}
\left[
\begin{array}{ccc} i&i&1\\ i&i&0\\ 1&0&1
\end{array}
\right]
 \frac{\left<\lambda'\mu' u=2 \left| [{\cal A}_{11}^\dagger(i)
\tilde{\cal A}_{20}(i)]^{(11)0} \right| \lambda \mu u=0
\right>}{\Delta C}
\right\}
\label{9}
\end{eqnarray}
where $\Delta C \equiv C(\lambda \mu)-C(\lambda' \mu')$, and the pairing
operators ${\cal A}$ are defined in Ref.\ \cite{wu93b}. Antisymmetrization
requires that $K_{1} + I_{1}$ and $K_{2} + I_{2}$ be even integers; therefore,
in Eq.\ (4) the allowed values for $K_{2}$ are $0$ and $2$ ($I_{2}=0$), and
$K_{1}$ can only be 1 since $I_1=1$. This means that for the $r=1$ term,
$\sum_{i}{ n}^{(11)0}_{i}e^{1}_{i}$ admixes the $u = 0$ and $u =2$ irreps by
changing a pair of particles in the symmetric $SU(3)$ representation $(\lambda,
\mu)=(2,0)$  into an antisymmetric representation $(0,1)$ through the
interaction ${\cal A}_{K_{1}1}^\dagger(i)\tilde{\cal A}_{K_{2}0}(i)$.  No other
terms associated with the single-particle energies have any influence on the
$SU(3)$ irreps of the $u=0$ space.

The methods of  Ref.\ \cite{che91b} may be used to evaluate the matrix
elements appearing in Eq.\ (4), and the preceding results may then be used to
calculate numerically the influence of realistic spherical single-particle
energies in the {\small FDSM}. Such work is in progress, but we now demonstrate
that perturbation theory may be employed to obtain an immediate estimate for
the limiting magnitude of $SU(3)$ symmetry-breaking caused by the
single-particle splitting in the $u=0$ representations. To second order in the
perturbation, the difference in energy with and without splitting is
\begin{equation}
\Delta E= \left|E-E(u=0)\right|=\frac{\Delta^{2}}
{E(\lambda',\mu',u'=2)-E(\lambda,\mu,u=0)}  ,
\label{10}
\end{equation} where $\Delta$ is the mixing matrix element associated with the
single-particle splitting [Eq.\ (4)]. As an upper limit, the matrix elements of
the pairing operators $[{\cal A}_{K_{1}1}^\dagger (i)\tilde{\cal
A}_{K_{2}0}(i)]^{(11)0}$ in Eq.\ (4) can be replaced by the diagonal matrix
element of the monopole pairing in the $(n_1,0)$ representation, which is
$\frac{1}{4}n_{i}(2\Omega_{i}/3-n_{i}+2 )$ (see Ref.\ \cite{wu93b}). The
particle number $n_i$ can be estimated as $(\Omega_{i}/\Omega_{1})n_{1}$, where
$n_1$ is the total number of particles in normal-parity levels and is around
$2\Omega_{1}/3$ for the most deformed nuclei.  Assuming that
$(\lambda,\mu)=(n_1,0)$ and
$\Delta C=C(n_1-2,1)-C(n_1,0)=-3n_1$,  the mixing matrix element $\Delta$ can
be evaluated from Eq.\ (4) if we make an assumption concerning the relative
phases of the two contributing terms.  Let us take as extreme cases the
assumption that the two matrix elements for $K_2=0$ and $K_2=2$ are either in
phase (a result denoted by $\Delta_>$) or out of phase (denoted by $\Delta_<$):
 $$
\Delta_< =-\frac{0.0787}{\Omega_1}\sum_{i}{e^{1}_{i}\Omega_{i}}
\left[
\begin{array}{ccc} i&i&1\\ i&i&0\\ 1&0&1
\end{array} \right]
\qquad
\Delta_> =-\frac{1.412}{\Omega_1}\sum_{i}{e^{1}_{i}\Omega_{i}}
\left[
\begin{array}{ccc} i&i&1\\ i&i&0\\ 1&0&1
\end{array} \right]
\label{11}  .
$$

The excitation energy $|E(\lambda',\mu',u'=2)-E(\lambda,\mu,u=0)|$ appearing in
Eq.\ (5) should be the energy required to break a pair (approximately the
pairing gap energy), plus the excitation energy due to the change of the
$SU(3)$ representation (which should be greater than the bandhead energy of the
$\gamma$ and $\beta$ bands). Thus its lower limit can be estimated as 2 MeV
for  normally deformed nuclei. The values of $e_j$ and the corresponding
$e^{r}_{i}$ in the $k-i$ basis for the experimental single-particle splittings
within a major shell are taken from \cite{dev83}. From these quantities, the
range of upper limits for $\Delta E$  versus the single-particle splitting can
be estimated;  the results are displayed in Fig.\ 1  for the 126--184 shell.
The quantities $\Delta E_<$ and $\Delta E_>$ are the energy shifts calculated
using $\Delta_<$ and $\Delta_>$, respectively.  Thus, the shaded areas in these
two figures represent  the range of expected {\em upper limits} for the
single-particle symmetry breaking.  The horizontal axis
$\rho=e_{j}/e_{j}(\mbox{exp})$ is a factor multiplying the spread of the
single-particle (s.~p.) energy scheme $\{e_{j}\}$ under consideration; hence
$\rho=1$ corresponds to the experimental s.~p.\ splitting, and  changing
$\rho$ corresponds to scaling the overall magnitude of the  splitting. It is
seen from Fig.\ 1 that the upper limit on the symmetry-breaking effect for the
$SU(3)$ dynamical symmetry caused by the s.~p.\ nondegeneracy  is very small
(from a fraction of an eV to several keV for the experimental spectra
appropriate to the  126--184 shells).  Only when the s.~p.\ energy splitting is
artifically large (say ten times larger than the splitting observed in a major
shell) will the effect of $SU(3)$ dynamical symmetry breaking be significant
and call into question the present perturbation theory analysis.  We reiterate
that the shaded region in Fig.\ 1 represents a range of estimated upper limits.
The average symmetry-breaking in realistic situations may be even smaller than
these estimates.

The small second-order energy shift justifies our use of perturbation theory
for the present estimates, and implies that the symmetry-breaking admixture in
the $SU_{3}$ wavefunction for $u=0$ configurations is perturbative in size.
Thus, the {\small FDSM} $SU_{3}$ wavefunction will remain essentially pure in
the  presence of the symmetry-breaking implied by a realistic spherical
single-particle spectrum. This suggests that the single-particle
symmetry-breaking terms will on average have only  small influence on other
observables such as transition rates and moments.

 The preceding discussion has been formulated in terms of the {\small FDSM} for
normal deformation, which assumes a single major shell of neutrons and protons
as a valence space, with effective interactions incorporating the influence of
the truncation.  An {\small FDSM} of superdeformations has been proposed
\cite{wu92b} that employs a valence space of mainly two oscillator shells  for
neutrons and protons.  An analysis similar to the present one may easily be
carried out for such spaces, but we can obtain an immediate estimate of the
influence of single-particle symmetry breaking for superdeformation from the
present results.  On the one hand, the summation over $i$ in Eq.\ (4) should
now run over two shells, which would increase the value of $\Delta$ by
approximately a factor of two; on the other hand, the pair degeneracy
$\Omega_{1}$ of the shells responsible for the $SU(3)$ symmetry will also
increase by about a factor of two in going to the superdeformed case. These two
effects approximately cancel each other, keeping the value of $\Delta$ nearly
the same for normal and superdeformation.  However, the energy denominator in
Eq.\ (5)  is at least a factor of two larger for superdeformed configurations
relative to the normally deformed case because of the enhanced collectivity.
{\em Thus, the effect of the single-particle symmetry breaking on the energies
is expected on general grounds to be even smaller for the superdeformed case
than  for the normally deformed case examined here.}  This analysis implies
that for superdeformed states, as for normally deformed states,  the collective
$SU_{3}$ wavefunction of the {\small FDSM} remains essentially pure in the
presence of realistic single-particle energy splitting.

This pronounced stability of the $SU(3)$ dynamical symmetry for the {\small
FDSM} is not a general property of fermion $SU(3)$ symmetries; it is a direct
consequence of the particular structure of the Ginocchio $S$--$D$ pairs from
which the $SU(3)$ symmetry of the {\small FDSM} is realized
\cite{gin80,wu87b,wu93b}. {\em The {\small\em FDSM} $SU(3)$ symmetry is defined
in the pseudoorbital ($k=1$) space}, with $U_k(3)\times U(\sum 2i+1)$ as its
higher symmetry ($U_k(3)$ is the unitary group associated with the $k$ degree
of freedom,  $U(\sum 2i+1)$ is the unitary group associated with the $i$
degrees of freedom). Therefore all the operators in the $k$-space must commute
with the $SU(3)$ invariants and there is no operator in the pseudospin space
that can break the $SU(3)$ symmetry. The only way in which the s.~p.\ terms can
break the $SU(3)$ symmetry is through the higher symmetry  $Sp(6)$ in which the
$SU(3)$ symmetry is embedded, because the total wavefunction is required to be
antisymmetric (i.\ e., the only allowed symmetry breaking is indirect, through
the Pauli effect).  This constraint can force a change in the $SU(3)$
representation if the irrep of the $i$ part of the wavefunction is changed. The
only operator in the s.~p.\ energy terms that can accomplish this is ${
n}^{(11)0}_{i}$. In other fermion theories, such as the Elliott model
\cite{ell58} or the pseudo-$SU(3)$ model \cite{ari69,hec69}, the $SU(3)$
symmetry is embedded in a much larger group; therefore, there are many
generators that could  break directly the $SU(3)$ symmetry and one generally
expects that the symmetry is  more susceptible to symmetry breaking by
single-particle energy terms.

It has been demonstrated  in the Ginocchio model that for the vibrational
symmetries of the {\small FDSM}, much of the effect of realistic
single-particle splitting can be absorbed by a renormalization of parameters
\cite{kir84}, leaving seniority as a reasonably good quantum number for
low-lying vibrational states.  We have shown here that single-particle symmetry
breaking for the  rotational symmetries of the {\small FDSM} has little
influence on the corresponding symmetry. {\em  Thus, the results of
\cite{kir84} and the present paper are strong evidence that  spherical
single-particle splitting leaves the collective aspects of all five dynamical
symmetries of the {\small FDSM} largely intact,   and  we may expect on general
grounds that collective properties  obtained in the {\small \em FDSM} symmetry
limits will survive the inclusion of realistic spherical single-particle
spectra.}

To summarize,  we have examined in this paper the effect of single-particle
energy nondegeneracies on the {\small FDSM} $SU(3)$ dynamical symmetry for
representations of zero heritage.  We find that the symmetry is essentially
preserved for any physically acceptable spectrum.  Thus, the phenomenological
successes of the {\small FDSM} $SU(3)$ model should survive in the limit of a
realistic spherical single-particle energy spectrum for the collective aspects
of both normal deformation and superdeformation.  This result demonstrates that
it is possible to construct  classes of fermion $SU(3)$ symmetries that are
virtually unperturbed even by large excursions in single-particle energies.  It
will be of interest to compare the present results with other fermion
$SU_{3}$ symmetries such as the Elliott Model and the pseudo-$SU_{3}$ model,
where one expects single-particle effects to have a larger influence on the
symmetries. It will also be of interest to enquire whether the present results
are unique to the {\small FDSM} and its underlying Ginocchio coupling scheme,
or whether there may exist additional classes of fermion symmetries having
unusual stability with respect to single-particle energy splitting.  Such
theories have not been widely discussed in nuclear structure, but are of
obvious interest for phenomena like identical bands that exhibit a particularly
large stability of nuclear properties with respect to changing particle number.

\vspace{10pt}
Nuclear physics research at Chung Yuan Christian University is supported by the
National Science Council of the {\small ROC.}  Nuclear physics research at
Drexel University is supported by the National Science Foundation. Theoretical
nuclear physics research at the University of Tennessee is supported by the
U.~S. Department of Energy through Contract No.\ DE--FG05--93ER40770. Oak Ridge
National Laboratory is managed by Martin Marietta Energy Systems, Inc.\ for the
U.~S. Department of Energy under Contract No.\ DE--AC05--84OR21400.

\noindent
Fig.\ 1 \hspace{5pt}
Range of upper limits for the energy shift (relative to the symmetry limit) in
$u=0$ irreps caused by realistic single-particle energies in the 126--184
shell.

\end{document}